  \magnification\magstep2
  \baselineskip = 0.5 true cm
  \parskip=0.5 true cm
                           
  \def\sa{\vskip 0.30 true cm}
  \def\sb{\vskip 0.60 true cm}
  \def\sc{\vskip 0.15 true cm}

  \pageno = 0   
  \vsize = 23   true cm
  \hsize = 16.3 true cm

  \def\gr{\hbox{\bf R}}
  
  \def\grn{\hbox{\bf N}}
  
  \def\grc{\hbox{\bf C}}

\rightline{LYCEN 9121}
\rightline{June 1991} 

\sc
\sb 

\centerline {\bf ON QUANTUM GROUPS AND THEIR POTENTIAL USE} 
\vskip 0.4 true cm                                       
\centerline {{\bf IN MATHEMATICAL CHEMISTRY} $^*$} 

\sa
\sb
\vskip 0.5 true cm

\centerline {Maurice Kibler$^1$ and Tidjani N\'egadi$^2$} 

\sa

\centerline {$^1$ Institut de Physique Nucl\'eaire de Lyon}
\centerline {IN2P3-CNRS et Universit\'e Claude Bernard}
\centerline {43 Boulevard du 11 Novembre 1918}
\centerline {F-69622 Villeurbanne Cedex, France}

\sa

\centerline {$^2$ Laboratoire de Physique Th\'eorique}
\centerline {Institut des Sciences Exactes} 
\centerline {Universit\'e d'Oran Es-S\'enia} 
\centerline {31100 Es-S\'enia, Alg\'erie}

\sa
\sa
\sb
\sb

\baselineskip = 0.7 true cm

\sa

\centerline {\bf Abstract}

\sb

The quantum algebra $su_q(2)$ is introduced as a deformation of 
the ordinary Lie algebra $su(2)$. This is achieved in a simple 
way by making use of $q$-bosons. In connection with the quantum 
algebra $su_q(2)$, we discuss the $q$-analogues of the harmonic 
oscillator and the angular momentum. We also introduce 
$q$-analogues of the hydrogen atom by means of a 
$q$-deformation of the Pauli equations and of the so-called 
Kustaanheimo-Stiefel transformation. 

\sa
\sa
\sa
\sb

\baselineskip = 0.5 true cm
\noindent $^*$ Paper published in 
Journal of Mathematical Chemistry 
{\bf 11}, 13-25 (1992). 
Paper written from a 
lecture presented (by M.~K.) at the ``IV International 
Conference on Mathematical and Computational Chemistry'', Bled 
(Yugoslavia), 3~-~7 June 1991. 

\vfill\eject
\baselineskip = 0.7 true cm

\centerline {\bf 1. Introduction}

A new algebraic structure, the 
structure of quantum group, has been developed since 1985 [1-3] 
and is still the subject of developments both in mathematics and 
theoretical physics. Such a structure, which is related to the 
structure of Hopf bi-algebra, takes its origin in various fields 
of theoretical physics (e.g., statistical mechanics, 
integrable systems, conformal field theory). 

The notion of quantum group is more easily approached through 
the one of quantum algebra. Loosely speaking, the latter notion 
corresponds to a deformation, depending on a certain parameter 
$q$, of a Lie algebra. Most of the applications of quantum 
algebras, of potential use for chemical physics, have been 
mainly devoted to the harmonic oscillator [4-8] and to coherent 
states [9,10].

It is the aim of this 
paper to briefly describe one of the simplest 
quantum groups, viz., the quantum group $SU_q(2)$, or rather 
its quantum algebra $su_q(2)$, and to 
underline its potential use in chemical physics. For this 
purpose, we examine in turn three dynamical systems connected 
with quantum groups : the $q$-deformed harmonic oscillator, 
the $q$-deformed angular momentum and the $q$-deformed hydrogen 
atom. These new systems, also referred to as $q$-analogues, 
reduce to the corresponding ordinary sytems in the limiting case $q=1$. 

The paper presents a review character as far as the 
$q$-analogues of the harmonic oscillator (in Section 2) and the 
angular momenta (in Section 3) are concerned. The discussion 
(in Section 3 and Appendix 2) 
about the relevance of the quantum algebra $so_q(3,2)$ for 
studying the $q$-analogues of spherical and hyperbolic angular 
momenta is new. The introduction (in Section 4) of 
$q$-analogues for the hydrogen atom is developed for the first 
time. No sophisticated mathematical pre-requisite is necessary to 
understand this self-contained article. 

\sb

\centerline {\bf 2. $q$-Analogue of the Harmonic Oscillator} 

We start with the usual Fock space 
$$
{\cal F} = \left \{ |n> \; \; : \; n \in \grn \right \}
\eqno (1)
$$
which is very familiar to the chemist.

Definition 1. Let us define the linear operators $a^+$, $a$ and 
$N$ on the vector space ${\cal F}$ by the relations
$$
a^+ \; |n> \; = {\sqrt {[n + 1]}} \; |n + 1 > \quad
a   \; |n> \; = {\sqrt {[n]    }} \; |n - 1 > \quad
N   \; |n> \; = n \; |n     > 
\eqno (2)
$$
with $a \; | 0> \; = 0$, where we use the notation
$$
[c] \; \equiv \; [c]_q \; = \; { {q^c - q^{-c}} \over {q - q^{-1}} }
\; = \; {{\sinh (c \ln q)} \over {\sinh (\ln q)}} \qquad c \in 
\grc \eqno (3)
$$
for a given $q$ in the field of complex numbers \grc. 

It is 
to be observed that in the limiting 
case $q=1$, we have simply $[c]=c$ 
so that $a^+$, $a$ and $N$ are (respectively) in this case the  
ordinary creation, annihilation and number operators 
encountered in various areas of theoretical chemistry and 
physics. In the case where $q \ne 1$, with $q$ not being a root of 
unity, the operators 
$a^+$, $a$ and $N$ defined by equations (2-3) are called $q$-deformed 
creation, annihilation and number operators, respectively. (In 
this case, the complex number $[c]$ defined by (3) is a 
$q$-deformed number; some algebraic relations 
satisfied by such $q$-deformed numbers are listed in Appendix 1.) 

Property 1. As a trivial property, we have
$$
(a)^{\dagger} = a^+ \qquad (N)^{\dagger} = N \qquad
[N,a^+] = a^+ \qquad [N,a] = -a 
\eqno (4)
$$
where $(X)^{\dagger}$ denotes the adjoint of the operator $X$ and
$[X,Y] \equiv [X,Y]_- = XY -YX$ the commutator of $X$ and $Y$.

Property 2. As a basic property, we can check that
$$
a a^+ = [N + 1]              \quad
a^+ a = [N]                  \quad
a a^+ - q^{-1} a^+ a = q^{N} \quad 
a a^+ - q      a^+ a = q^{-N}
\eqno (5)
$$
where we use the abbreviation 
$$
[X] \; \equiv \; [X]_q \; = \; { {q^X - q^{-X}} \over {q - q^{-1}} }
\; = \; {{\sinh (X \ln q)} \over {\sinh (\ln q)}} 
\qquad X \in {\cal F} \eqno (6)
$$
which parallels for operators the defining relation (3) for 
numbers.

The set $\left \{ a, a^+\right \}$ satisfying (4-6) is a set of 
$q$-bosons as originally defined by Macfarlane [4] and 
Biedenharn [5] (see also Refs.~[6,7]). 
From equation (5), it is clear that the 
operators $a$ and $a^+$ reduce to ordinary bosons 
in the limiting case $q=1$.

We are now in a position to introduce a $q$-deformed harmonic 
oscillator. The literature on this subject is now abundant and 
the reader may consult, for example, Refs.~[4-10] for further details.

Definition 2. From the $q$-deformed creation and annihilation 
operators $a$ and $a^+$, let us define the operators
$$
p_x \; = \; i  \; \sqrt{ {\hbar \mu \omega} \over 2 }  \; (a^+ - a)
\qquad 
  x \; =       \; \sqrt{ \hbar \over {2 \mu \omega} }  \; (a^+ + a)
\eqno (7)
$$
acting on ${\cal F}$, 
where $\hbar$, $\mu$ and $\omega$ have their usual meaning in 
the context of the (ordinary) harmonic oscillator.

Equation (7) defines $q$-deformed momentum and position operators
$p_x$ and $x$, respectively, and bears the same form as for the ordinary 
creation and annihilation operators corresponding to the 
limiting case $q=1$.

Property 3. The commutator of the $q$-deformed 
operators $x$ and $p_x$ is
$$
[x, p_x] \; = \; i \hbar \; ([N+1] - [N]) \eqno (8)
$$
which reduces to the ordinary value $i\hbar$ in the limiting 
case $q=1$.

In terms of eigenvalues, equation (8) can be rewritten as
$$
[x, p_x] \; = \; i \hbar \; 
{\cosh [(n + {1 \over 2}) \ln q] \over \cosh ({1 \over 2} \ln q)}
\eqno (9)
$$
when $q \ne 1$. Thus, we may think of a $q$-deformed 
uncertainty principle: the right-hand side of (9)
increases with $n$ (i.e., with the energy, see equation (11) below) and is 
minimum as well as $n$-independent in the limiting case $q = 1$ 
[5].

Definition 3. We define the self-adjoint operator $H$ on ${\cal F}$
by
$$
H = {1 \over {2\mu}} \, {p_x}^2 + {1 \over 2} \, \mu \, \omega^2 
\, x^2 = {1 \over 2} \, (a^+a \, + \, aa^+) \, \hbar \, \omega
= {1 \over 2} \left( [N]  +  [N+1] \right) \, \hbar \, \omega 
\eqno (10)
$$
in terms of the $q$-deformed operators previously defined.

In the limiting case $q=1$, the operator $H$ is nothing but the 
Hamiltonian for a one-dimensional harmonic oscillator. 
Following Macfarlane [4] and Biedenharn [5], we take 
equation (10) as the defining relation for a $q$-deformed 
one-dimensional harmonic oscillator. The case of a $q$-deformed 
$d$-dimensional, with $d \ge 2$, (isotropic or anisotropic)
harmonic oscillator can be handled from a superposition of 
one-dimensional $q$-deformed oscillators. 

Property 4. The spectrum of $H$ is given by
$$
E \; \equiv \; E_n \; = \; {1 \over 2} \; ([n] + [n+1]) \; \hbar \; \omega
\; = \; [2]_{q^{1 \over 2}} \; {1 \over 2} \; [n + {1 \over 2}] 
\; \hbar \; \omega \qquad n \in \grn \eqno (11)
$$
and is discrete.

This spectrum turns out to be a deformation of the one for the 
ordinary one-dimensional harmonic oscillator corresponding to 
the limiting case $q=1$. The levels are shifted (except the 
ground level) when we pass from $q=1$ to $q \ne 1$ : 
the levels are not uniformly spaced.

\sb 

\centerline {\bf 3. $q$-Analogues of Angular Momenta} 

We now continue with the Hilbert space
$$
{\cal E} \, = \, \left\{ |jm> \ \; : \; 2j \in \grn, \ m = -j(1)j \right\} 
\eqno (12)
$$
spanned by the common eigenvectors of the $z$-component and the 
square of a generalized angular momentum.

Definition 4. We define the operators  
operators $a_+$,    $a_+^+$, 
          $a_-$ and $a_-^+$ on the vector space ${\cal E}$ by the relations
$$
\eqalign{
a_+ \; |jm> \; = \; & {\sqrt{[j + m]}}\; |j - {1\over 2}, m - {1\over 2}>\cr
a_+^+ \; |jm> \; = \; & {\sqrt{[j + m+1]}}\; |j + {1\over 2}, m + {1\over 2}>\cr
a_- \; |jm> \; = \; & {\sqrt{[j - m]}}\; |j - {1\over 2}, m + {1\over 2}>\cr
a_-^+ \; |jm> \; = \; & {\sqrt{[j - m + 1]}}\; |j + {1\over 2}, m - {1\over 2}>
\cr
} \eqno (13)
$$
where the numbers of the type $[c]$ are given by (3).

In the limiting case $q=1$, equation (13) gives back the 
defining relations used by Schwinger [11] in his 
(Jordan-Schwinger) approach to angular momentum (see also Ref.~[12]). 
By introducing 
$$
n_1 = j + m \qquad n_2 = j - m \qquad n_1 \in \grn \qquad n_2 \in \grn
\eqno (14)
$$  
and
$$
|jm> \, \equiv \, |j + m, j - m> \, = \, |n_1n_2> \; \; \in \; 
{\cal F}_1 \otimes {\cal F}_2 \eqno (15)
$$
equation (13) can be rewritten in the form
$$\eqalign{
  a_+   \; |n_1n_2> \; = \; &{\sqrt {[n_1]}}\; |n_1 - 1, n_2>\cr
  a_+^+ \; |n_1n_2> \; = \; &{\sqrt {[n_1 + 1]}}\; |n_1 + 1, n_2>\cr
  a_-   \; |n_1n_2> \; = \; &{\sqrt {[n_2]}}\; |n_1, n_2 - 1>\cr
  a_-^+ \; |n_1n_2> \; = \; &{\sqrt {[n_2 + 1]}}\; |n_1, n_2 + 1>\cr
  } \eqno (16)
$$  
Therefore, the sets $\left\{ a_+, a_+^+ \right\}$ 
                   and $\left\{ a_-, a_-^+ \right\}$ are two commuting sets of 
$q$-bosons. More precisely, we can prove that
$$
a_+a^+_+ \; - \; q^{-1} a^+_+ a_+ \; = \; q^{N_1} \qquad 
a_-a^+_- \; - \; q^{-1} a^+_- a_- \; = \; q^{N_2}$$
$$
  [a_+, a_-]     \; = \; 
  [a^+_+, a^+_-] \; = \;
  [a_+, a^+_-]   \; = \;
  [a^+_+, a_-]   \; = \; 0
\eqno (17)
$$
with 
$$N_1 |n_1n_2 > \; = \; n_1 |n_1n_2> \qquad 
  N_2 |n_1n_2 > \; = \; n_2 |n_1n_2 > \eqno (18)$$
defining the number operators $N_1$ and $N_2$.

Definition 5. Let us consider the operators 
$$
J_- \; = \; a^+_- a_+ \qquad 
J_3 \; = \; {1 \over 2} \left( N_1 - N_2 \right) \qquad
J_+ \; = \; a^+_+ a_- 
\eqno (19)
$$
defined in terms of $q$-bosons.

Property 5. The action of the linear operators $J_-$, $J_3$ and $J_+$ 
on the space ${\cal E}$ is described by
$$
\eqalign{
  J_- \; |jm > \; = \; &{\sqrt {[j + m] \; [j - m + 1]}} \; |j, m-1 >\cr  
  J_3 \; |jm > \; = \; &m \; |jm >\cr
  J_+ \; |jm > \; = \; &{\sqrt {[j - m] \; [j + m + 1]}} \; |j, m+1 >\cr
}
\eqno (20)
$$
a result that follows from (13) and (19).

The operators $J_-$ and $J_+$ are clearly shift operators for 
the quantum number $m$. The operators $J_-$, 
$J_3 = (J_3)^{\dagger}$ and 
$J_+ = (J_-)^{\dagger}$ 
reduce to ordinary spherical angular momentum operators in the 
limiting case $q=1$. The latter assertion is evident from (20) 
or even directly from (19).

At this stage, the quantum algebra $su_q(2)$ can be introduced, 
in a pedestrian way, from equations (19) and (20) as a 
deformation of the ordinary Lie algebra of the special unitary 
group $SU(2)$. In this regard, we have the following property.

Property 6. The commutators of the $q$-deformed spherical angular 
momentum operators $J_-$, $J_3$ and $J_+$ are 
$$
[J_3, J_-] \; = \; - \; J_- \qquad 
[J_3, J_+] \; = \; + \; J_+ \qquad 
[J_+, J_-] \; = \; [2J_3]
\eqno (21)
$$
which reduce to the familiar expressions known in angular 
momentum theory in the limiting case $q=1$.

Equation (21) is at the root of the definition of the quantum 
algebra $su_q(2)$. Roughly speaking, this algebra is spanned by 
any set ${J_-, J_3, J_+}$ of three operators satisfying (21) where 
we recognize familiar commutators except for the third one. 
The notion of invariant operator also exists for 
quantum algebras. In this connection, we can verify that the 
operator
$$
J^2 \; = \; {1 \over 2} \; 
(J_+J_- + J_-J_+) + {{[2]} \over {2}} \; [J_3]^2
\eqno (22)
$$
is a Casimir operator in the sense that it commutes with each of 
the generators $J_-$, $J_3$ and $J_+$ of the quantum algebra 
$su_q(2)$. It can be proved that the eigenvalues of the 
hermitian operator $J^2$ are $[j] \, [j+1]$ with $2j \in \grn$, 
a result compatible with the well-known one corresponding to the 
limiting case $q=1$. 

Definition 6. We now introduce the operators
$$
K_- \; = \; a_+a_- \qquad
K_3 \; = \; {1 \over 2} \; (N_1 + N_2 + 1) \qquad
K_+ \; = \; a^+_+ a^+_- 
\eqno (23)
$$
which are indeed $q$-deformed hyperbolic angular momentum 
operators. 

Property 7. The action of the operators $K_-$, $K_3$ and $K_+$ 
on the space ${\cal E}$ is described by
$$
\eqalign{
  K_- \; |jm > \; = \; &{\sqrt {[j-m] \, [j+m]}}     \; |j-1, m >\cr
  K_3 \; |jm > \; = \; &(j+{1 \over 2}) \; |jm >\cr
  K_+ \; |jm > \; = \; &{\sqrt {[j-m+1] \, [j+m+1]}} \; |j+1, m >\cr
}
\eqno (24)
$$
a result to be compared to (20).

The operators $K_-$ and $K_+$ behave like shift operators for 
the quantum number $j$. The operators $K_-$, 
$K_3 = (K_3)^{\dagger}$ and 
$K_+ = (K_-)^{\dagger}$ 
reduce to ordinary hyperbolic angular momentum operators in the 
limiting case $q=1$ [11,12]. From equation (24), we expect that 
they generate the quantum algebra $su_q(1,1)$, a result which 
is trivial when $q=1$.

Property 8. The commutators of the $q$-deformed hyperbolic angular 
momentum operators $K_-$, $K_3$ and $K_+$ are 
$$
[K_3, K_-] \; = \; - \; K_- \qquad 
[K_3, K_+] \; = \; + \; K_+ \qquad
[K_+, K_-] \; = \; - \; [2K_3]
\eqno (25)
$$
which characterizes the quantum algebra $su_q(1,1)$.

Equations (20) and (21), on one hand, and equations (24) and 
(25), on the other, can serve to develop the theory of 
$q$-deformed spherical and hyperbolic angular momenta. This 
theory involves the $q$-deformation of coupling 
(Clebsch-Gordan) coefficients and recoupling (Racah and Wigner) 
coefficients, projection operators, etc.~and shall not be 
described here (see, among numerous papers, Ref.~[13]). In the 
limiting case $q=1$, the Wigner-Racah algebra of $SU(2)$, in an 
$SU(2) \supset U(1)$ basis, plays a considerable r\^ole in this 
theory ; in this case, the Lie algebra of the de Sitter group $SO(3,2)$ is the 
natural framework for studying the Wigner-Racah algebra of 
$SU(2)$. We devote the rest of this section to some basic 
elements indicating the relevance of the quantum algebra 
$so_q(3,2)$ when $q \ne 1$.

Definition 7. We define the operators
$$
k^+_+ \; = \; - \; a^+_+ a^+_+ \qquad 
k^+_- \; =      \; a^+_- a^+_- \qquad
k^-_- \; = \; - \; a_+ a_+     \qquad
k^-_+ \; =      \; a_- a_-
\eqno (26)
$$
in terms of $q$-bosons.

Property 9. The action of the operators 
 $k^+_+$, 
 $k^+_-$, 
 $k^-_-$ and 
 $k^-_+$ 
on the space ${\cal E}$ is described by
$$
\eqalign{
  k^+_+ \; |jm > \; = \; &- {\sqrt {[j+m+1] \, [j+m+2]}} \; |j+1, m+1 >\cr
  k^+_- \; |jm > \; = \; &  {\sqrt {[j-m+1] \, [j-m+2]}} \; |j+1, m-1 >\cr
  k^-_- \; |jm > \; = \; &- {\sqrt {[j+m-1] \, [j+m]}}   \; |j-1, m-1 >\cr
  k^-_+ \; |jm > \; = \; &  {\sqrt {[j-m-1] \, [j-m]}}   \; |j-1, m+1 >\cr
}
\eqno (27)
$$
so that they act like mixed step operators for the quantum 
numbers $j$ and $m$. 

Some further properties, 
of interest for the quantum algebra $so_q(3,2)$,
of the operators of type $J$, $K$ and $k$ 
are relegated on Appendix 2.

\sb 

\centerline {\bf 4. $q$-Analogue of the Hydrogen Atom}

We now consider an (ordinary) hydrogenlike atom in 3 dimensions 
with reduced mass $\mu$ and nuclear charge $Ze$. We deal here only 
with the discrete spectrum of this (Coulomb) dynamical system, i.e., with 
negative energies $E$.

According to Pauli [14], the Coulomb system can be described in 
an operator form by the equations (see also Ref.~[15])
$$
A^2 - B^2 \; = \; 0 \qquad \quad 
E \, \left( 2A^2 + 2B^2 + \hbar^2 \right) \; 
= \; - {1 \over 2} \; \mu \; Z^2 \; e^4
\eqno (28)
$$
In equation (28), the operators $A^2 = \sum_i A_i^2$ and
                                $B^2 = \sum_i B_i^2$
stand for the Casimir operators of the Lie algebras $asu(2)$ and 
$bsu(2)$, of type $su(2)$, 
spanned by $\left\{ A_i \ : \ i = 1,2,3 \right\}$ and 
           $\left\{ B_i \ : \ i = 1,2,3 \right\}$, 
respectively, where 
$$
{A}_i \; = \; {1\over 2} ({L}_i + {N}_i) \qquad \quad
{B}_i \; = \; {1\over 2} ({L}_i - {N}_i) \qquad \quad
{N}_i \; = \; \sqrt{ {-\mu} \over {2E} } \; {M}_i
\eqno (29)
$$
In equation (29), $L_i$ ($i = 1,2,3$) and 
                  $M_i$ ($i = 1,2,3$) denote the components 
of the angular momentum operator and the Laplace-Runge-Lenz-Pauli vector 
operator, respectively. 

The transition from the ordinary hydrogen atom to a $q$-deformed 
hydrogen atom can be achieved by passing from the (direct sum) Lie algebra 
$asu  (2) \oplus bsu  (2) \sim so(4)$ to the quantum algebra 
$asu_q(2) \oplus bsu_q(2)$. The application of this deformation 
to equation (28) leads to the $q$-analogue of the 
hydrogen(like) atom whose energy spectrum is given by 
$$
E \; \equiv \; E_j \; = \; { {1} \over {4[j] [j+1] + 1} } \; E_{0} 
\qquad \quad 2j \in \grn 
\eqno (30)
$$
where 
$$
E_{0} \; = \; - \; {1\over 2} \; { {\mu \; Z^2 \; e^4} \over {\hbar^2} }
\eqno (31)
$$
is the energy of the ground state.

The $q$-deformed atom thus defined has the same ground 
state energy as the ordinary atom. The other states are shifted 
when passing from $q=1$ to $q \ne 1$. The whole (discrete) 
spectrum of the $q$-deformed hydrogen atom exhibits the same degeneracy 
as the ordinary one. Of course, the $q$-deformed spectrum 
coincides with the ordinary one when $q$ goes to 1.

To close this section, we should mention there are other ways 
to define a $q$-analogue of the hydrogen atom which do not lead 
to the spectrum (30-31). In this respect, by using the 
Kustaanheimo-Stiefel transformation (see Ref.~[15]), we are 
left with a $q$-deformed hydrogen atom characterized by the 
discrete spectrum 
$$
E \equiv E_{n_1n_2n_3n_4} \; = 
                          \; { {16} \over {\nu(n_1n_2n_3n_4)^2} } \; E_{0}
$$
$$
\nu(n_1n_2n_3n_4) 
\; = \; \sum_{i = 1}^{4} \; [n_i] + [n_i + 1] \qquad n_i \in \grn 
\qquad (i = 1,2,3,4) 
\eqno (32)
$$
Equation (32) can be derived (i) by transforming the 
three-dimensional hydrogen atom into a four-dimensional 
isotropic harmonic oscillator by means of the 
Kustaanheimo-Stiefel transformation [15], (ii) by passing from 
the latter oscillator to its $q$-analogue and (iii) by invoking 
the ``inverse'' Kustaanheimo-Stiefel transformation. The result 
(32) thus obtained constitutes an alternative to (30).

\sb 

\centerline {\bf 5. Closing Remarks}

We have concentrated in the present paper on $q$-deformations 
of three dynamical systems (harmonic oscillator, angular 
momentum and hydrogen atom) largely used in physical chemistry. 
The $q$-deformed dynamical systems have been introduced in 
connection with the quantum algebra $su_q(2)$ which turns out 
to be a deformation of $su(2)$ characterized by the deformation 
parameter $q$. 

We have seen that the parameter $q$ enters the 
(energy) spectra for the $q$-analogues of the considered dynamical
systems. There is no universal significance of the parameter 
$q$. However, in view of the fact that the limiting case $q=1$ 
gives back the usual spectra, the deformation parameter $q$ 
might be considered as a fine structure parameter (like a 
curvature constant), to be obtained from a fitting procedure, 
for describing small effects. In addition, it may happen in 
some situations that it is worth to consider $q$ as a 
completely free parameter with values far from 1 leading to new 
models [16]. 

We have experienced that the correspondence between the 
hydrogen atom and its $q$-analogue is not one-to-one. (This is 
indeed a general problem we face when dealing with 
$q$-analogues.) As a remedy, the use of the $q$-derivative leading 
to a $q$-deformed Schr\"odinger equation might be interesting.
Also, the use of sets of noncommuting $q$-bosons might be 
appropriate to ensure $su_q(2)$ covariance. 

\sb 

\centerline {\bf Acknowledgments}

One of the authors (M.~K.) thanks Y. Saint-Aubin for communicating 
his lecture notes (Ref.~[17]) on quantum groups. He is 
grateful to J. Katriel and S.~L. Woronowicz for interesting 
discussions.

\sb 

\centerline {\bf Appendix 1}

In this appendix we give some formulas useful for dealing with 
$q$-deformed numbers $[c]$ when $c$ are real numbers or integers.

From equation (3), we easily get
$$
\lim_{q \to 1} \; [c]_q \; = \; c
$$
$$
[-c]_q \; = \; - [c]_q \qquad [c]_{1 \over q} \; = \; [c]_q
$$
$$
[c]_q \geq c \quad \hbox {for} \quad c > 1
$$
Furthermore, the following relations 
$$
[a + b] = [a] \, q^b + q^{-a} \, [b]
$$
$$
[a + 1] \, [b + 1] - [a] \, [b] = [a + b + 1]
$$
$$
[a] \, [b+c] = [a+c] \, [b] + [a-b] \, [c]
$$
$$
[a]^2 - [b]^2 = [a-b] \, [a+b]
$$
hold for any (real) numbers $a$, $b$ and $c$.

In the case where $n$ is a positive integer, we have
$$
[n] = {\sum_{i=(1-n)(2)(n-1)}} 
  q^i = q^{n-1} + q^{n-3} +  ...  + q^{-n+3} + q^{-n+1}
\quad n \in \grn - \left \{ 0 \right \}
$$
and we can define the factorial of $[n]$ as
$$
[n]! \; = \; [n] \, [n-1] \, ... \, [1] \qquad n \in \grn \qquad [0]! = 1
$$
As illustrative examples, we have 
$$[0] = 0 \qquad [1] = 1 \qquad [2] = q^{-1}+q$$
$$[3] = q^{-2} + 1 + q^2 \qquad [4] = q^{-3} + q^{-1} + q + q^3$$
and
$$
[2] \, [2] = [1] + [3] \qquad
[2] \, [3] = [2] + [4] \qquad
[3] \, [3] = [1] + [3] + [5]
$$
which is reminiscent of the addition rule for angular momenta.

In the case where $q$ is a root of unity, we have
$$
q = \exp ( {i2\pi {{k_1}\over {k_2}}} ) \qquad
  k_1 \in \grn \qquad k_2 \in \grn 
$$
$$
[c] = {{\sin(2\pi{{k_1}\over {k_2}}c)}\over
       {\sin(2\pi{{k_1}\over {k_2}})}}
$$
For instance, 
$$
k_1 = 1 \quad k_2 = 4 \quad \Rightarrow \quad q = i = {\sqrt {-1}}
\quad \Longrightarrow \quad [0] = [2] = [4] = ... = 0 
$$
so that $[c] = 0$ can occur for $c \ne 0$.

\vfill\eject 

\centerline {\bf Appendix 2}

It is a simple matter of calculation to determine the 
commutation relations between the 10 operators of type $J$, $K$ 
and $k$ defined in Section 3. We list in the following only the 
nonvanishing commutators. The arrow indicates the limit when 
$q$ goes to 1.
  
\noindent {Nonvanishing $[k,k]$ matrix elements :}
$$
[k^+_+, k^-_-] \; = \; - \; [2K_3 + 2J_3 - 1] \; 
                       - \; [2K_3 + 2J_3 + 1] \; 
        \rightarrow \; - \; 4(K_3 + J_3)
$$
$$
[k^+_-, k^-_+] \; = \; - \; [2K_3 - 2J_3 - 1] \; 
                       - \; [2K_3 - 2J_3 + 1] \; 
        \rightarrow \; - \; 4(K_3 - J_3)
$$
\noindent {Nonvanishing $[J,K]$ matrix elements :}
$$\matrix{
  [J_+, K_+] & = 
 &k^+_+ ( [ K_3 - J_3 - {1\over 2} ] &-&
          [ K_3 - J_3 + {1\over 2} ] )
 &\rightarrow &- \; k^+_+\cr
  [J_+, K_-] & = 
 &k_+^- ( [ K_3 + J_3 - {1\over 2} ] &-&
          [ K_3 + J_3 + {1\over 2} ] )
 &\rightarrow &- \; k^-_+\cr
  [J_-, K_+] & = 
 &k^+_- ( [ K_3 + J_3 + {1\over 2} ] &-&
          [ K_3 + J_3 - {1\over 2} ] )
 &\rightarrow &+ \; k^+_-\cr
  [J_-, K_-] & =
 &k^-_- ( [ K_3 - J_3 + {1\over 2} ] &-&
          [ K_3 - J_3 - {1\over 2} ] )
 &\rightarrow &+ \; k^-_-\cr
}$$
\noindent {Nonvanishing $[J,k]$ matrix elements :}
$$
  [J_3, k^+_+] \;  = \;   k^+_+ \quad 
  [J_3, k^+_-] \;  = \; - k^+_- \quad 
  [J_3, k^-_-] \;  = \; - k^-_- \quad
  [J_3, k_+^-] \;  = \;   k^-_+ 
$$
$$\matrix{
    [J_+,k^+_-] & = 
 &K_+ ( [ K_3 - J_3 + {3 \over 2} ] &-&
        [ K_3 - J_3 - {1 \over 2} ] )
 &\rightarrow &+ \; 2K_+\cr
    [J_+,k^-_-] & = 
 &K_- ( [ K_3 + J_3 + {1 \over 2} ] &-&
        [ K_3 + J_3 - {3 \over 2} ] )
 &\rightarrow &+ \; 2K_-\cr
    [J_-,k^+_+] & = 
 &K_+ ( [ K_3 + J_3 - {1 \over 2} ] &-&
        [ K_3 + J_3 + {3 \over 2} ] )
 &\rightarrow &- \; 2K_+\cr
    [J_-,k^-_+] & =
 &K_- ( [ K_3 - J_3 - {3 \over 2} ] &-&
        [ K_3 - J_3 + {1 \over 2} ] )
 &\rightarrow &- \; 2K_-\cr
}$$
\noindent {Nonvanishing $[K,k]$ matrix elements :}
$$
  [K_3, k^+_+] =   k^+_+ \quad
  [K_3, k^+_-] =   k^+_- \quad 
  [K_3, k^-_-] = - k^-_- \quad 
  [K_3, k_+^-] = - k^-_+ 
$$
$$
\matrix{
    [K_+,k^-_-] & = 
 &J_- ( [ K_3 + J_3 + {1 \over 2} ] &-&
        [ K_3 + J_3 - {3 \over 2} ] )
 &\rightarrow &+ \; 2J_-\cr
    [K_+,k^-_+] & = 
 &J_+ ( [ K_3 - J_3 - {3 \over 2} ] &-&
        [ K_3 - J_3 + {1 \over 2} ] )
 &\rightarrow &- \; 2J_+\cr
    [K_-,k^+_+] & = 
 &J_+ ( [ K_3 + J_3 - {1 \over 2} ] &-&
        [ K_3 + J_3 + {3 \over 2} ] )
 &\rightarrow &- \; 2J_+\cr
    [K_-,k^+_-] & =
 &J_- ( [ K_3 - J_3 + {3 \over 2} ] &-&
        [ K_3 - J_3 - {1 \over 2} ] )
 &\rightarrow &+ \; 2J_-\cr
}
$$
From the commutation relations in this appendix and in Section 
3, we recover that the set $\left \{ J,K,k \right \}$ spans the 
(10-dimensional) noncompact Lie algebra $so(3,2) \sim sp(4, \gr)$ 
in the limiting case $q=1$ [12].

\vfill\eject 

\centerline {\bf References}

\baselineskip = 0.55 true cm

 \noindent [1] V.G. Drinfel'd {\it Sov. Math. Dokl.} {\bf 32} 254 (1985).

 \noindent [2] M. Jimbo {\it Lett. Math. Phys.} {\bf 10} 63 (1985).

 \noindent [3] S.L. Woronowicz {\it Comm. Math. Phys.} {\bf 111} 613 (1987).
 
 \noindent [4] A.J. Macfarlane {\it J. Phys. A: Math. Gen.} 
{\bf 22} 4581 (1989).

 \noindent [5] L.C. Biedenharn {\it J. Phys. A: Math. Gen.} 
{\bf 22} L873 (1989).
 
 \noindent [6] C.-P. Sun and H.-C. Fu {\it J. Phys. A: Math. Gen.} 
{\bf 22} L983 (1989).
 
 \noindent [7] P.P. Kulish and N.Yu. Reshetikhin {\it Lett. 
Math. Phys.} {\bf 18} 143 (1989).

 \noindent [8] M. Nomura {\it J. Phys. Soc. Jpn.} {\bf 59} 2345 
(1990).

 \noindent [9] C. Quesne {\it Phys. Lett.} {\bf 153A} 303 (1991).

 \noindent [10] J. Katriel and A.I. Solomon {\it J. Phys. A: Math. Gen.} 
{\bf 24} 2093 (1991). 

 \noindent [11] J. Schwinger, On angular momentum, {\it Report} 
U.S. AEC NYO-3071 (1952). (Published in: {\it Quantum Theory of 
Angular Momentum}, eds. L.C. Biedenharn and H. van Dam (New York: 
Academic, 1965).) 

 \noindent [12] M. Kibler and G. Grenet {\it J. Math. Phys.} 
{\bf 21} 422 (1980).

 \noindent [13] Yu.I. Kharitonov, Yu.F. Smirnov and V.N. Tolstoy, 
 Method of the projection operators and $q$-analog of the 
 quantum angular momentum theory, {\it Reports} 1607 
 and 1636, Institut of Nuclear Physics, Academy of Sciences of 
 the USSR, Leningrad (1990).

 \noindent [14] W. Pauli {\it Z. Phys.} {\bf 36} 336 (1926). 

 \noindent [15] M. Kibler and T. N\'egadi {\it Lett. Nuovo Cimento} 
{\bf 37} 225 (1983); {\it J. Phys. A: Math. Gen.} {\bf 16} 4265 
(1983); {\it Phys. Rev.} A {\bf 29} 2891 (1984).

 \noindent [16] J.A. Tuszy\'nski and M. Kibler (work in progress).

 \noindent [17] Y. Saint-Aubin, Quantum groups and their 
 application to conformal quantum field theories, {\it Report} CRM-1663, 
 Universit\'e de Montr\'eal (1990).

\bye